# Effect of Ba and Zr co-substitution on dielectric and magnetoelectric properties of BiFeO$_3$ multiferroics

Satya N.Tripathy[1,2]*, Dhiren K. Pradhan[3], S. Sen[4], Braja G. Mishra[5], R. Palai[3], J. F. Scott[6], R. S. Katiyar [2] and Dillip K. Pradhan[1]*

[1]Department of Physics and Astronomy, National Institute of Technology, Rourkela-769008, India.

[2]Department of Physics, Koneru Lakshmaiah Education Foundation, Deemed to be University, Hyderabad-500075, Telangana, India

[3]Department of Physics and Institute for Functional Nanomaterials, University of Puerto Rico, San Juan, PR 00936, USA.

[4]Sensor and Actuator Division, Central Glass and Ceramics Research Institute, Kolkata-700032, India.

[5]Department of Chemistry, National Institute of Technology, Rourkela-769008, India.

[6]Department of Physics, Cavendish Laboratory, University of Cambridge, Cambridge, CB3 OHE, United Kingdom

## ABSTRACT

We report the effect of Ba and Zr co-substitution on structural, dielectric, magnetic and magnetoelectric properties of BiFeO$_3$ multiferroics. Polycrystalline nanoceramic samples of Bi$_{1-x}$Ba$_x$Fe$_{1-y}$Zr$_y$O$_3$ ($0.0 \leq x = y \leq 0.1$) have been synthesized by auto-combustion method. Rietveld refinement result of X-ray diffraction data indicate a contraction of unit cell volume with increase in $x$ and $y$. Field emission scanning electron micrographs show densely populated grains without any voids or defects with well-defined grain boundary and decrease in grain size with increasing $x$ and $y$. The result of dielectric measurement display anomalies, which are amplified and the temperature at which dielectric anomalies observed decreases with increasing composition. A cross-over from anti-ferromagnetism to weak ferromagnetism has been observed at $x = y = 0.1$ with enhanced magnetization as compared to pure BiFeO$_3$. Ferroelectric properties of the sample have been studied using PE loop measurement. Magneto-dielectric and magneto impedance spectroscopy reveal the existence of intrinsic magneto-electric coupling at room temperature.

Author for correspondence E-mail: satyanarayantripathy@gmail.com and dillip.pradhan79@gmail.com

# I. INTRODUCTION

Multiferroic materials are scientifically and technologically important because of their fascinating fundamental physics[1-5] and potential applications[6-8] in novel multifunctional devices. So far, a very limited number of materials display room temperature (RT) multiferroic properties (i.e., magnetic and electric ordering in single phase) owing to chemical incompatibility and mutual exclusiveness of these orderings[9-10]. $BiFeO_3$ (BFO) is a prototype single-phase multiferroic material with high ferroelectric transition temperature ($T_C$ ~830 °C) and antiferromagnetic (AFM) Néel temperature ($T_N$ ~370 °C)[9]. However, there are some issues still needing to be resolved for the better understanding of magnetoelectric coupling and device applications[4-5]. One of the major issues associated with the material is the synthesis of stoichiometric BFO. The narrow range of phase stabilization of $BiFeO_3$ (from $Bi_2O_3$-$Fe_2O_3$ phase diagram) lead to the appearance of secondary impurity phases, such as $Bi25FeO40$ and $Bi_2Fe_4O_9$ [3-4]. Therefore, synthesis of single-phase BFO is a challenging problem. The presence of impurity phases gives rise to poor/artificial ME coupling which are of extrinsic in nature [4-5,9] resulting small spontaneous polarization with high electrical leakage current [9]. So far, most of the reported multiferroics show weak magnetization and low magnetoelectric coupling at RT, which is the main barrier for multifunctional device applications. This is due to the occurrence of canted long-range spin cycloid modulation with period of 620–640 Å. Hence, the linear magneto-electric effect and remnant magnetization collapse on a macroscopic scale[11-12]. The ultimate application of multiferroic materials for functional devices requires strong magnetoelectric coupling at RT. Recent theoretical and experimental studies suggest that the improvement of magnetoelectric coupling of BFO at RT can be achieved by adopting the following approaches: (a) suppression of the spin cycloid, and/or (b) driving the ferroic phase transition temperatures $T_C$ or $T_N$ towards room temperature [13-14]. These approaches can be accomplished by the following strategies: (a) suitable chemical substitutions; (b) preparation of nanoparticles smaller than the dimension of the spin cycloid; (c) application of high magnetic fields; (d) fabrication of epitaxial thin films[11-12, 15-21]. Chemical substitutions have been successfully established as an appropriate methodology to improve the multiferroic properties of BFO[22]. This may be due to the possibility of driving the crystal symmetry of BFO near to a morphotropic phase boundary (MPB) and improving the piezoelectric, magnetic and magnetoelectric properties[22]. Several reports have shown enhancement in multiferroic properties of BFO by synthesizing nanoparticles smaller than spin cycloid, due to the size effect[11,20].

In order to enhance magnetoelectric coupling, solid solution of BFO with rare earth manganites (RMnO$_3$) and ferroelectric perovskites (ABO$_3$) have been widely investigated [23,9]. Recently Khomchenko *et al.* reported the crystal structure and enhanced multiferroic properties of diamagnetic (i.e., $Ca^{2+}$, $Sr^{2+}$, $Pb^{2+}$ and $Ba^{2+}$) substituted BFO[24]. Wei *et al.* observed the phenomenon of cross-over from antiferromagnetic to ferromagnetic in BFO due to nonmagnetic $Zr^{4+}$ substitution[25]. Mukherjee *et al.* studied the influence of Zr- substitutions on structure, electrical and ferroelectric properties of BFO thin films. They reported that the maximum extend of Zr- substitutions will be below $x = 0.15$ without the formation of secondary phases [26, 27]. Catlan *et al.* have investigated the effect of $Ca^{2+}$ substitution on the improved magnetoelectric coupling of BFO[22]. Kawae *et al.* showed the improved ferroelectric and leakage current behaviour in Mn- and Ti-modified $BiFeO_3$ thin films[28]. It has been found that doping of $Zr^{4+}$ at the $Fe^{3+}$ site in BFO introduces Bi-vacancies, decreases the Fe-O-Fe bond angle, and enhances the super exchange interaction and magnetic ordering due to the tilting of oxygen octahedral[25]. However, the effect of co-doping of non-magnetic ions on both the A- and B-sites of BFO has not been investigated. In the present work, we doped BFO with $Ba^{2+}$ and $Zr^{4+}$ at the $Bi^{3+}$ and $Fe^{3+}$ sites respectively, to compensate the cation deficiency and avoid Bi-vacancy due to $Zr^{4+}$ doping.

We provide a detailed description of $Ba^{2+}$ and $Zr^{4+}$ co-substitution effect on the structural, microstructural, dielectric, magnetic, calorimetric and magneto-dielectric properties of BFO nanoceramics. A deeper understanding of structural parameters and its correlation with magnetic and magnetoelectric properties is also presented.

## II. EXPERIMENTAL

Nanoceramic of $Bi_{1-x}Ba_xFe_{1-y}Zr_yO_3$ ($0.0 \leq x = y \leq 0.1$) were prepared by auto-combustion method using analytical grade chemicals of $Fe(NO_3)_3 \cdot 9H_2O$, $Bi(NO_3)_3 \cdot 5H_2O$, $Ba(NO_3)_2$, $ZrO(NO_3)_2 \cdot H_2O$ as starting materials (oxidizers) and urea as fuel. Details of auto-combustion synthesis procedure are available in our previous works [29, 30]. For $x = y = 0.0$, the *combustion residue* was calcined at an optimized temperature of 550 °C for 3 h in air, whereas for $0.025 \leq x = y \leq 0.1$ the combustion residues were calcined at 780 °C for 3h. The calcined powders were pressed in form of cylindrical pellets using hydraulic press with pressure $8 \times 10^7$ kg.m$^{-2}$ using polyvinyl alcohol (PVA) as binder. The pellets were sintered at 780 °C for 12 h. The phase formation of the compounds was checked by X-ray powder diffractometer (Philips Analytical-PW 3040) at RT. The X-ray diffraction (XRD) data were collected at very slow scan of 2°/min with a step size of 0.02° in a wide range of Bragg angles 2θ (20° ≤2θ ≤ 80°) using Cu-K$_{α1}$ radiation (λ = 1.5405 Å). Crystallographic analyses of the materials were carried out by Rietveld technique using FULLPROF package[31]. The peak shapes and background were modelled using a pseudo-Voigt function and six-coefficient polynomial function, respectively. The zero correction, scale factor, peak shape parameters, background, unit cell parameters, atomic positions, thermal parameters, and half-width parameters (U, V and W) were varied during refinement process while the occupancies of all the atoms were kept fixed. The microstructural properties of the samples were estimated and analysed using X-ray line profile analysis by BREADTH software[32]. For electrical characterization, the pellets were polished, painted with silver electrodes and dried at 150 °C for 3 h to remove the moisture. The dielectric parameters (*i.e.,* capacitance, loss tangent, impedance, phase angle) were measured in a wide frequency range (*i.e.,* 100 Hz to 1 MHz) using a computer-controlled impedance analyzer PSM-1735 in the temperature range 25-400 °C using a home-made sample holder. Field emission scanning electron microscope (Nova Nano SEM 450) was used to determine grain size, and sample matrix defects. Calorimetric measurements were carried out with a Mettler-Toledo DSC apparatus equipped with liquid nitrogen cooling accessory and a HSS8 ceramic sensor (a heat flux sensor with 120 thermocouples). The magneto-dielectric measurements were performed using a vibrating sample magnetometer (VSM Lakeshore model 142A) and a HIOKI 3532-50 LCR Hister at RT. Magnetization hysteresis measurements (M-H hysteresis loop) were carried out using a vibrating sample magnetometer up to a maximum field of ±2 Tesla at RT and the temperature dependent M-H hysteresis loop measurement were carried out using physical properties measurement system (PPMS-Quantum Design) up to a maximum field of ±8 Tesla.

## III. Results and Discussion

### A. Structural Study

Fig.1 (a) depicts the Rietveld refinement result of X-ray diffraction data of $Bi_{1-x}Ba_xFe_{1-y}Zr_yO_3$ ($0.0 \leq x = y \leq 0.1$) at RT. Above $x$ and $y \geq 0.10$, we observed secondary phases, which is consistent with 10% Zr modified BFO[26,27,33]. So, we have restricted our investigations for $0.0 \leq x = y \leq 0.1$. However the presence of small amount of impurity phases (i.e, $Bi_2Fe_4O_9$ which is 2 to 4 % of the parent phase calculated from the integrated intensity analysis) cannot be avoided under this synthesis conditions for higher concentration of Ba and Zr co-substitutions. Similar observations of appearance of impurity phases have been reported for higher concentration of Zr-substituted BFO[27]. The experimental data points are represented as symbol (+) and fitted data are shown as solid line and difference is at the bottom of the plot (Fig. 1 (a)). There are three ions ($Bi^{3+}/ Ba^{2+}$, $Fe^{3+}/ Zr^{4+}$, and $O^{2-}$) in the asymmetric unit of the rhombohedral perovskite structure. We have adopted the representations of atomic positions for hexagonal unit cell of $R3c$ space group as per Megaw and Darlington[34]. The coordinates of all the atoms in the asymmetric unit cell of the $R3c$ space group can also be written as a function of the atomic displacement parameters $s, t, d$, and $e$ : $Bi^{3+}/ Ba^{2+}$ $(0, 0, \frac{1}{4} + s)$, $Fe^{3+}/ Zr^{4+}$ $(0, 0, t)$, $O^{2-}$ $(\frac{1}{6} - 2e - 2d, \frac{1}{3} - 4d, \frac{1}{12})$. The parameters "$s$" and "$t$" describe the polar displacement of cations $Bi^{3+}/ Ba^{2+}$ and $Fe^{3+}/ Zr^{4+}$ along polar axis of the rhombohedral $[111]_{rh}$ or the hexagonal $[001]_h$ axis. The displacement parameter "$e$" of oxygen $O^{2-}$ from its ideal position is related to the tilt angle (ω) of anti-phase rotation of the oxygen octahedra about the rhombohedral $[111]_{rh}$ direction through the expression $\omega = \tan^{-1}(4e3^{1/2})$. The parameter "$d$" is related to the distortion of the $BO_6$ (B: $Fe^{3+}$/ $Zr^{4+}$) octahedra. The $z$- coordinate of the oxygen ions is anchored at $z = 1/12$ [34]. For $x = 0.0$ a best fitting is obtained between calculated and observed XRD spectra (goodness of fit: $\chi^2 = 1.59$) with refined lattice parameters of unit cell $a = 5.5779$ (09) Å and $c = 13.8710$ (23) Å, which agree with reported values[18, 30]. The XRD spectra of all Ba-Zr modified samples are analogous to the parent BFO ($R3c$ space group). This is evidenced by existence of characteristic superlattice reflection (113) around 38 degree of the rhombohedral $R3c$ phase. Therefore, we have selected the $R3c$ space group to refine the crystal structure for all compositions. The relative intensity of the super-lattice peak (113) reflection decreases with increasing $x$ (Fig.1 (b)). Since the (113) reflection arises from anti-ferrodistortive tilting of the octahedra, this gives rise to new oxygen planes leading to low intensity[16]. The Wyckoff notations and atomic positions used in Rietveld refinement for the $R3c$ space groups are listed in Table.1. Fig. 1(c) shows the variation of the normalized lattice parameters (i.e., $a_H^N = a_{Hex}/\sqrt{2}$, $c_H^N = c_{Hex}/\sqrt{12}$) and ($c_H^N/a_H^N$) of the $R3c$ phase with composition. It shows that on increasing Ba-Zr content, normalized lattice parameters and pseudo-tetragonality ($c_H^N/a_H^N$) decreases. So, the Ba and Zr co-substation does not give rise to structural transitions but leads to significant structural distortion and change in lattice parameters. Similar observations have been reported for Zr modified BFO[27, 33]. The variations of $s$, $t$, $d$, $e$, and $\omega$ atomic displacement parameters with composition have been represented in Table. 2. It is to be noted that $s$, $t$, and $\omega$ parameters decrease on increasing composition $x$. Furthermore, in order to demonstrate the effect of peak broadening due to small crystallite size qualitatively and quantitatively, we have adopted the double-Voigt method using a BREATH package[32], which is similar to the Warren-Averbach method. Fig 1(d) shows the volume-weighted domain size distribution function $P_V(L)$ as a function of crystallite size (L) for $0.0 \leq x =$ $y \leq 0.1$. It is evident from the figure that the true crystallite size of the samples varies between 100 Å to 600Å.

Fig. 2 shows the field emission scanning electron micrographs for $x = y = 0.0$, 0.05 and 0.1. As can be seen, the grains are densely populated with well-defined grain boundary without any voids or defects. BiFeO$_3$ ($x = y =0.0$) shows highly non-uniform grain size distributions ranging from ~300 nm to ~3 µm, whereas the BZ-doped samples show average grain size of 30-60 nm.

**B. Dielectric and Calorimetric Study**

For multiferroic materials, temperature-dependent dielectric properties have been widely investigated to provide the signature of magnetoelectric coupling. Fig. 3(a) compares the dielectric permittivity $\varepsilon(T)$ as a function of temperature of Bi$_{1-x}$Ba$_x$Fe$_{1-y}$Zr$_y$O$_3$ system for $0.0 \leq x = y \leq 0.1$ at 100 kHz. For $x = y = 0.0$ two anomalies were observed around 215 °C and 365°C (small kink), as shown in Fig 3 (c). The observed anomaly around 365 °C is attributed to the antiferromagnetic phase transition ($T_N$), whereas the anomaly around 215 °C ($T_{ME}$) due to higher-order - magnetoelectric coupling[30]. This could be due to the suppression of the spatially modulated spin structure below the magnetic ordering temperature[35, 36]. With increasing composition, both the anomalies gradually become more prominent and show increase in the magnitude of dielectric permittivity (Fig 3 (d) for $x = y = 0.05$ as representative). It has been observed that the temperatures corresponding to the anomalies decreases with increase in composition. The temperature at which we observed the dielectric anomaly corresponds to $T_N$ except $x = y = 0.1$, which will be discussed in the section magnetic properties. The appearance of theses anomalies suggesting the signature of magneto-electric coupling in the systems. But for $x = y = 0.1$, a new anomaly is seen around 56 °C along with the two anomalies mentioned above.

Here $Ba^{2+}$ and $Zr^{4+}$ have been substituted at the Bi and Fe sites of BFO. This aliovalent substitutions lead to valence fluctuations due to the different oxidation state of Ba and Zr. Due to the presence of different oxidation states, there will be different kind of ME effect between the different cation elements via oxygen, which gives rise to competition between different magnetic ordering (which will be discussed in magnetic properties) leading to a magnetic frustration in the system [37]. So the appearance of the new peaks could be due to the magnetic frustration of paramagnetic and weakly ferromagnetic orderings and/or could be due to the fluctuation of cations as the ionization potential of $Zr^{4+}$ (34 eV) is quite high compared to that of $Ba^{2+}$ (10 eV). The temperature dependent loss tangent (tan δ) also showed an anomaly, where the dielectric anomaly has been observed (Fig. 3b), further supporting the signature of magneto-electric coupling. Since dielectric measurements are done at high frequency, to avoid interfacial effects, the observed dielectric anomalies related to the bulk properties and could be the signature of ME coupling[38]. As can be seen from Figs 3(a) and (b), we observed negative dielectric permittivity at high temperature. However, we cannot correlate this effect to metal-insulator transition as the contribution of negative dielectric permittivity is not due to the (metallic) resistor: In fact, it is an inductance, not a resistance, that is involved. It is known from the famous work of Jonscher on 'universal dielectric response'[39, 40] and also from the Feynman lecture series book,[41] that all capacitors (even the simplest) have an ac response that is partly inductive. Under the conditions of our experiments, this induction becomes non-negligible, and hence the apparent dielectric response can extrapolate to negative values; but this is just a phase shift, not a 'negative capacitance. The negative permittivity is also central to the nature of resonant interaction of structure surface with radiation[42]. Doping ions (metal and nonmetal) into polymer results in negative dielectric permittivity due to the induction effect at optical frequency (above plasma

frequency)$^{42}$. In our case, frequency is very low and it is inadequate to create resonant condition. We did not observe any negative dielectric permittivity in the frequency dependent dielectric measurement (Fig. 8) at same frequency range.

Fig. 4 shows the variation of electrical polarization (P) with the applied electric field (E) with different dc electric fields of $Bi_{1-x}Ba_xFe_{1-y}Zr_yO_3$ for $x = y = 0.1$ as representative at room temperature. The ferroelectric measurements were carried out after poling the sintered pellet please put the value at 1200 V for 12 hours. It has been observed from the graph that, no saturation polarization could be achieved in P-E which may be due to the existence of leakage current in the samples. The coercive field and remnant polarization increase with increase of electric field. This behavior suggest the ferroelectric nature of the samples. The gap in the negative polarization axis is due to the temporary memory that decays away in a few seconds. Moreover, we have carried out a calorimetric study to examine existence of dielectric anomalies in temperature dependent dielectric plot. The endothermic peaks takes place within experimental uncertainty around the same temperature as observed in the dielectric study. The DSC endothermic peak corresponding to dielectric anomaly decreases with increase in $x = y$ (see Fig. 5). The observed anomalies in temperature dependent dielectric properties are attributed to intrinsic magneto-electric coupling in the samples.

## C. Magnetic Properties

Figs. 6(a) and (b) show the magnetic hysteresis loop of $Bi_{1-x}Ba_xFe_{1-y}Zr_yO_3$ ($0.0 \leq x = y \leq 0.1$) measured using VSM. Pure BFO shows a linear M-H loop without any spontaneous magnetization at RT, a typical behavior of antiferromagnetic ordering. This can be described in terms of the presence of canted G-type AFM spiral spin modulation in BFO$^9$. With increasing composition up to $x$ and $y = 0.05$ M-H loop has similar behavior as that of pure BFO, but improved magnetic properties were observed for $x = y = 0.075$ and $x = y = 0.1$. For $x = y = 0.1$ magnetic hysteresis loop is observed with $H_c = 100$ mT and $M_r = 0.04$ emu/g. However, no saturation has been observed up to field of 2 T. The enhanced magnetization could be due to the suppression of the spin cycloid present in BFO induced either from co-doping and/or grain size smaller than the spin cycloid period. When $Fe^{3+}$ ions are replaced by $Zr^{4+}$ ions in the AFM ordering of $Fe^{3+}$sub-lattice, the balance between two adjacent anti-parallel spins of $Fe^{3+}$ gets perturbed. Eventually a ferromagnetic super-exchange interaction via oxygen, instead of AFM coupling, is preferred due to the straightening out of the Fe-O-Zr bond angle [25, 43]. Hence $Zr^{4+}$ substitution offers a unidirectional alignment of spins in the system, leading to ferromagnetic behavior. From structural study it is evident that lattice strain rises with increasing composition, since the ionic radii of $Ba^{2+}$ (1.49 Å) and $Zr^{4+}$ (0.86 Å) are larger than the ionic radii of $Bi^{3+}$ (1.17 Å) and $Fe^{3+}$ (0.78 Å). This facilitates the decrease in distance between $Fe^{3+}$ ions and oxygen and leads to stronger super-exchange interaction, resulting enhanced magnetic interaction and magnetization[25, 43]. Moreover, as mentioned above, synthesis of nanoparticle diameters less than the length of the spin cycloid could further enhance the magnetic properties, due to suppression of cycloid spin structure, which agrees with our X-ray line profile analysis. As mentioned above, we have observed the small amount of impurity secondary phases ($Bi_2Fe_4O_9$) for higher concentration of Ba and Zr co-substitutions. Cheng *et. al* [44] in pure BFO and Lin *et. al.* [45] in pure BFO and 10% La-doped BFO observed an anomaly in the magnetization in the magnetization vs temperature (*M* vs. *T*) curve due to the presence of the impurity phase $Bi_2Fe_4O_9$. $Bi_2Fe_4O_9$ is paramagnetic at room temperature and undergoes a transition to antiferromagnetic state near 264 K. So anomaly observed around 260 K by Cheng *et. al* [44] and Lin et. al.[45] is due to the antiferromagnetic phase transition of $Bi_2Fe_4O_9$. As the magnetic data presented here is at room temperature and $Bi_2Fe_4O_9$ is paramagnetic at RT, we expect the good hysteresis loop is due the Ba and Zr co-substitutions of BFO but not due to the impurity phases.

For the better understanding magnetic ordering in $x = y = 0.1$ sample at high magnetic field (up to 8 T) and high temperature (up to 800 K), we used PPMS for the measurement of MH loop. Figure 7 shows the MH loop and temperature dependent magnetization using PPMS. As can be seen, we observed magnetic hysteresis loop at RT consistent with VSM measurements. The observation of no saturation up to 8 T implied the presence of some paramagnetic phase in our sample, which is consistent with the presence of small impurity phase as detected by XRD. The magnetic measurement of BFO based multiferroic sample is quite tricky. This is because of difficulty in the synthesis of single phase BFO. Impurity phases like $Fe_3O_4$ (Tc= 575 °C), $\alpha Fe_2O_3$ (Tc= 675 °C) and $\gamma Fe_2O_3$ (Tc= 600 °C) are usually observed in BFO, which significantly affects the magnetic properties of BFO. As can been from the magnetization vs. temperature measurement (Fig. 7), we observed only one magnetic transition at 405 °C (onset 307 °C) for sample $x = y = 0.1$ and the magnetization is almost zero after 405 °C, which clearly rules out the presence of magnetic impurity phases. We do observed a very small amount of paramagnetic phase ($Bi_2Fe_4O_9$) but this phase will not contribute to the magnetic moment. Defects and valence fluctuations can also contribute to the magnetic moment. However, since our sample is almost single phase, we believe that the contribution from defect and valence fluctuation could be minimal. The dielectric anomaly observed in dielectric permittivity at around 230 °C for the sample $x = y = 0.1$ cannot be related to the magnetic transition temperature. The temperature-dependent magnetization (Fig. 7) for sample $x = y = 0.1$ shows the magnetic transition temperature at round 405 °C (onset of transition is around 307 °C). However, the dielectric anomaly observed for sample with x = y < 0.1 at around 330 °C, coincidently close to the $T_N$ of BFO (370 °C).

**D. Magneto-electric Coupling**

In multiferroic systems, magnetoelectric (ME) coupling means studies of switching of electrical polarization (**P**) by the application of magnetic field or equivalently switching of magnetization (**M**) by the application of electric field. In order to study the ME coupling between two ferroic parameters (**M** and **P**), we have to measure the change in magnetization by the application of electric field or vice versa. Many multiferroics are tend to be poor insulator so that they are unable to sustain the high electric field necessary to switch the magnetization. Thus determination of magneto-electric coupling using this process is quite difficult.[30, 46] Alternatively, the ME coupling can also be studied through the magneto-dielectric effect, where the dielectric and electrical properties will be studied under the application of magnetic field and vice versa. The dielectric constant relates to the index of the refraction (n) of the material n = $(\mu\varepsilon)^{1/2}$, where µ and ε are the permeability and permittivity of the material respectively. The tuning of refractive index by the application of a magnetic field. In order to provide deeper understanding of magneto-electric coupling in $Bi_{1-x}Ba_xFe_{1-x}Zr_xO_3$ system, we have studied the magneto-dielectric (MD) properties (i.e., measurement of dielectric properties as a function of magnetic field and frequency). The presence of magnetoelectric coupling in multiferroics has been established by the observation of two criteria[46]: (i) the observation of an anomaly in the temperature-dependent dielectric permittivity near the magnetic phase transition temperature; (ii) the changes in the capacitance with the application of magnetic field[44,47]. To establish the second criterion, we have analysed the frequency-dependent dielectric permittivity for $x = y = 0.025$, $x =$ $y = 0.05$, $x = y = 0.075$ and $x = y = 0.1$ at RT (Fig. 8) as a function of magnetic fields ($0.0 \leq H \leq 2$ T: $\Delta H = 0.5$ Tesla). As can be seen, the dielectric permittivity decreases with increase in magnetic field, indicating a negative magneto-dielectric effect. The product of spin-pair correlation of neighbouring spins and the coupling constant decide the negative or positive sign of the magneto-dielectric effect[48, 49]. Multiferroic (ferromagnetic-ferroelectric) materials under a magnetic field will produce magnetostrictive strain and if there is coupling with the ferroelectric (piezoelectric) ordering, it can induce an electric field[48, 49]. This directly results in magnetic field-induced change in dielectric constant as observed. However, in sintered polycrystalline samples such a magneto-dielectric effect may not necessarily result from multiferroicity of the bulk contributions, as it may have significant contributions from space-charge effects at the interfacial layers (i.e., grain boundaries, grain-electrode interfaces) of different resistivity[17]. However multiferroic systems with the mismatch of work function at the electrode-dielectric interfaces can cause a strong change in the dielectric properties on the application of magnetic field, which are not related to intrinsic properties of the materials[38]. These extrinsic effects can be ruled out by magneto-impedance (MI) measurement. Therefore, we have carried out magneto-impedance spectroscopic measurement to explore the intrinsic multiferroic effect[46]. Using this magneto-impedance spectroscopic technique, the contributions of interfacial phenomena can be separated out from the bulk contributions. The observed bulk capacitance is intrinsic in nature without any external effect. Fig.9 compares the RT complex impedance spectra (i.e., Nyquist plot: -Im Z vs. Re Z) for $x = y = 0.075$ at different magnetic fields. The Nyquist plot is characterized by the presence of two overlapping semi-circular arcs with their centres below the real axis. The high-frequency semicircle is attributed to the bulk (grain) property of the material, whereas the low-frequency arcs are related to the presence of grain boundary effect. The intercept of the semi-circular arcs on the real axis gives rise to bulk and grain boundary resistance of the materials[50, 51]. The diameter of semicircle increases with increasing magnetic field that implies the increase of bulk resistance (see Table 3). The assignment of the two semi-circular arcs to the electrical response due to grain interior and grain boundary is consistent with the brick-layer model for a polycrystalline material. For ideal Debye-like ideal response, the equivalent circuit comprises of a parallel combination of a resistor and capacitor. The combined impedance and modulus spectroscopic plot (-Z" and M" vs. frequency) are plotted in log-log scale in order to highlight the departure from the ideality and to suggest the equivalent circuit model as shown in inset of Fig. 9 [50, 51]. It is observed from the combined impedance and modulus spectroscopic that the two peak maxima are not frequency-coincident, suggesting the departure from ideal Debye behaviour and justifying the presence of a constant phase element. So the high frequency semicircular arc can be modelled to an equivalent circuit of parallel combination of a resistance (bulk), capacitance (bulk) along with a constant phase element, whereas the low frequency semicircular arc can be modelled for parallel combination of a resistance (grain boundary) and a capacitance (grain boundary) as shown in Fig. 9. Both the equivalent circuit corresponding to grain and grain boundary contribution are connected in series for fitting the impedance data. The impedance data (symbols) have been fitted (solid line) with the model proposed above via commercially available software ZSIMP-WIN Version 3. The bulk resistance and bulk capacitance obtained after fitting the impedance data are tabulated in Tables 3 and 4. Here, it is further intriguing to mention that the bulk permittivity (calculated from impedance data) decreases with increase in magnetic field (see Table. 4). The magnitude of maximum magneto-capacitance $MC\% = \frac{\varepsilon(H,T) - \varepsilon(0,T)}{\varepsilon(0,T)}$ increased up to -6.5 % at $H = 2$ T for $x$ and $y = 0.1$. Therefore, we conclude that the observed dielectric response of these materials under magnetic field is an indicative of presence of magnetoelectric effect.

## IV. SUMMARY AND CONCLUSIONS

In conclusion, solid solutions of $Bi_{1-x}Ba_xFe_{1-y}Zr_yO_3$ ($0.0 \leq x = y \leq 0.1$) were successfully synthesized sing an auto-combustion method. Rietveld refinement results showed that all the samples have rhombohedral crystal structure with *R3c* space group. FESEM showed the decrease in grain size with increasing composition (*x* and *y*). Temperature variation of dielectric permittivity showed an anomaly around the magnetic transition temperature, indicating the existence of the magnetoelectric coupling which is confirmed by calorimetric characterization. Magnetic hysteresis loop measurements at RT showed enhancement in magnetization with weak ferromagnetic behaviour for higher concentration of *x*. The existence of PE and MH hysteresis loop suggest the multiferroic properties of the materials. Frequency dependent magneto-dielectric measurements and Magneto Impedance studies showed the existence of magneto-electric coupling at RT. Hence, we conclude that Ba-Zr co-substitution plays an important role in improving the magneto-electric properties of BFO.

**ACKNOWLEDGMENTS**: This work was partially supported by DST fast track Project No: SR/FTP/PS-16/2009. The work at UPR was supported by National Science Foundation (DMR 1410869) and Institute for Functional Nanomaterials (IFN). We also acknowledge Dr. R. Schmidt (Universidad Complutense de Madrid, Spain) for fruitful suggestion on magneto-impedance spectroscopy. SNT also thankful to Prof. Marian Paulch (Institute of Physics, University of Silesia) for providing calorimetric measurements.

Table.1 Atomic positions for hexagonal unit cell of $R3c$ space group as per Megaw and

| Composition ($x = y$) | Space group | Atom | Site | x | y | z |
|---|---|---|---|---|---|---|
| $0.0 \leq x$ and $y \leq 0.1$ | $R3c$ | $Bi^{3+}/Ba^{2+}$ | 6a | 0.0 | 0.0 | 0.2972 (15) |
| | | $Fe^{3+}/Zr^{4+}$ | 6a | 0.0 | 0.0 | 0.0193 (16) |
| | | $O^{2-}$ | 18b | 0.2318 (43) | 0.3373 (45) | 1/12 (Fixed) |

Darlington used in Rietveld refinement. Standard deviations for $x = y = 0.025$ are in the parenthesis.

Table. 2 Compositional dependence of atomic displacement parameters $s$, $t$, $d$, $e$ and $\omega$ obtained

| $x = y$ | s | t | d | e | $\omega$ |
|---|---|---|---|---|---|
| 0.0 | 0.0472 (06) | 0.0193 (16) | -0.0010 (1E-5) | -0.0316 (06) | 12.34 |
| 0.025 | 0.0458 (08) | 0.0189 (17) | -0.0029 (8E-5) | -0.03528 (02) | 13.72 |
| 0.05 | 0.0385 (05) | 0.0127 (19) | 0.0006 (2E-5) | -0.03333 (03) | 12.99 |
| 0.075 | 0.0371 (07) | 0.0150 (23) | 0.0004 (3E-5) | -0.02905 (02) | 11.37 |
| 0.1 | 0.0402 (05) | 0.0232 (24) | -0.0013 (8E-5) | -0.02537 (03) | 10.0 |

from Rietveld refinement. Standard deviations for are in the parenthesis.

**Table. 3** Bulk resistance obtained from fitting of magnetic field dependent complex impedance plot at RT for different value of x =y. The error presented here is due to fitting of the experimental data using circuit model.

| Bulk Resistance in Ohm [Error less ≤ 3%] | | | | |
|---|---|---|---|---|
| Field (Tesla) | $x = y = 0.025$ $10^8$ | $x = y = 0.05$ $10^5$ | $x = y = 0.075$ $10^4$ | $x = y = 0.1$ $10^4$ |
| 0.0 | 1.594 | 3.01 | 4.614 | 3.240 |
| 0.5 | 1.608 | 3.04 | 4.626 | 3.841 |
| 1.0 | 1.744 | 3.038 | 4.636 | 4.021 |
| 1.5 | 1.774 | 3.092 | 4.653 | 4.106 |
| 2.0 | 1.775 | 3.0945 | 4.678 | 4.159 |

**Table. 4** Bulk capacitance obtained from fitting of magnetic field dependent complex impedance plot at RT for different value of $x = y$. The error presented here is due to fitting of the experimental data using circuit model.

| Bulk Capacitance ($10^{-11}$) in F [Error less ≤ 3%] | | | | |
|---|---|---|---|---|
| Field (Tesla) | $x = y = 0.025$ | $x = y = 0.05$ | $x = y = 0.075$ | $x = y = 0.1$ |
| 0.0 | 2.327 | 2.064 | 2.153 | 3.788 |
| 0.5 | 2.315 | 2.097 | 2.101 | 3.536 |
| 1.0 | 2.314 | 2.096 | 2.09 | 3.562 |
| 1.5 | 2.313 | 1.943 | 2.091 | 3.559 |
| 2.0 | 2.313 | 1.942 | 2.087 | 3.545 |

**Figure Captions**

**Figure.1** (Colour online) **(a)** Rietveld refinement result of XRD pattern of $Bi_{1-x}Ba_xFe_{1-y}Zr_yO_3$ for $0.0 \leq x$ and $y \leq 0.1$ at RT, **(b)** XRD pattern expanded in $36° \leq 2\theta \leq 42°$ indicating (113) super reflection of $R3c$ phase **(c)** Compositional dependence of normalized lattice parameters (*i.e.*, $a_H^N = a_{Hex}/\sqrt{2}$, $c_H^N = c_{Hex}/\sqrt{12}$) and ($c_H^N / a_H^N$) and, **(d)** Volume-weighted domain size distribution function, $P_v(L)$ as a function of crystallite size (L).

**Figure.2** (Colour online) FESEM micrographs of of $Bi_{1-x}Ba_xFe_{1-y}Zr_yO_3$ for **(a)** $x = y = 0.0$ **(b)** $x = y = 0.05$ **(c)** $x = y = 0.1$.

**Figure.3** (Colour online) Temperature dependent dielectric parameters of $Bi_{1-x}Ba_xFe_{1-y}Zr_yO_3$ for $0.0 \leq x$ and $y \leq 0.1$ **(a)** $\varepsilon_r$ vs. T at 100 kHz, **(b)** tan $\delta$ vs. T at 100 kHz. Temperature dependent dielectric permittivity at selected frequencies between 10 kHz -1 MHz for **(c)** $x = y = 0.0$ (Inset-temperature dependence of tan $\delta$ for $x = y = 0.0$) and **(d)** $x = y = 0.05$ (Inset-temperature dependence of tan $\delta$ for $x = y = 0.05$).

**Figure. 4** (Colour online) Room temperature P–E loops of $Bi_{1-x}Ba_xFe_{1-y}Zr_yO_3$ for $x = y = 0.1$ as representative.

**Figure. 5** (Colour online) DSC thermogram of **(a)** $x = y = 0.0$, **(b)** $x = y = 0.05$, **(c)** $x = y = 0.075$ and **(d)** $x = y = 0.1$.

**Figure.6** (Colour online) **(a)** M-H hysteresis loops of $Bi_{1-x}Ba_xFe_{1-y}Zr_yO_3$ **(a)** for $0.00 \leq x$ and $y \leq 0.075$ (Inset- magnification of MH loop at lower fields) **(b)** for $x = y = 0.1$ at RT.

**Figure.7** (Colour online) **(a)** M-H hysteresis loops of $Bi_{1-x}Ba_xFe_{1-y}Zr_yO_3$ for $x = y = 0.1$ at different temperatures, **(b)** Temperature dependence of magnetization of $Bi_{1-x}Ba_xFe_{1-y}Zr_yO_3$ for $x$

= $y$ = 0.1 measured with zero field cooling (ZFC) and field cooling (FC) with applied field of 1000 Oe.

**Figure. 8** (Colour online) Frequency dependent behaviour of dielectric permittivity for of (a) $x = y = 0.0$, (b) $x = y = 0.05$, (c) $x = y = 0.075$ and (d) $x = y = 0.1$ **a**t different applied magnetic fields ($0 \leq H \leq 2$ T*:* $\Delta H = 0.5$ T).

**Figure. 9** (Colour online) Complex impedance spectra at different applied magnetic fields ($0 \leq H \leq 2$ T*:* $\Delta H = 0.5$ T) for $x = y = 0.075$ at RT. Solid symbols represent experimental data while solid lines represent simulated curve using the equivalent circuit illustrated (Inset- Equivalent circuit model with combined impedance and modulus spectroscopic plot for H = 2 Tesla).

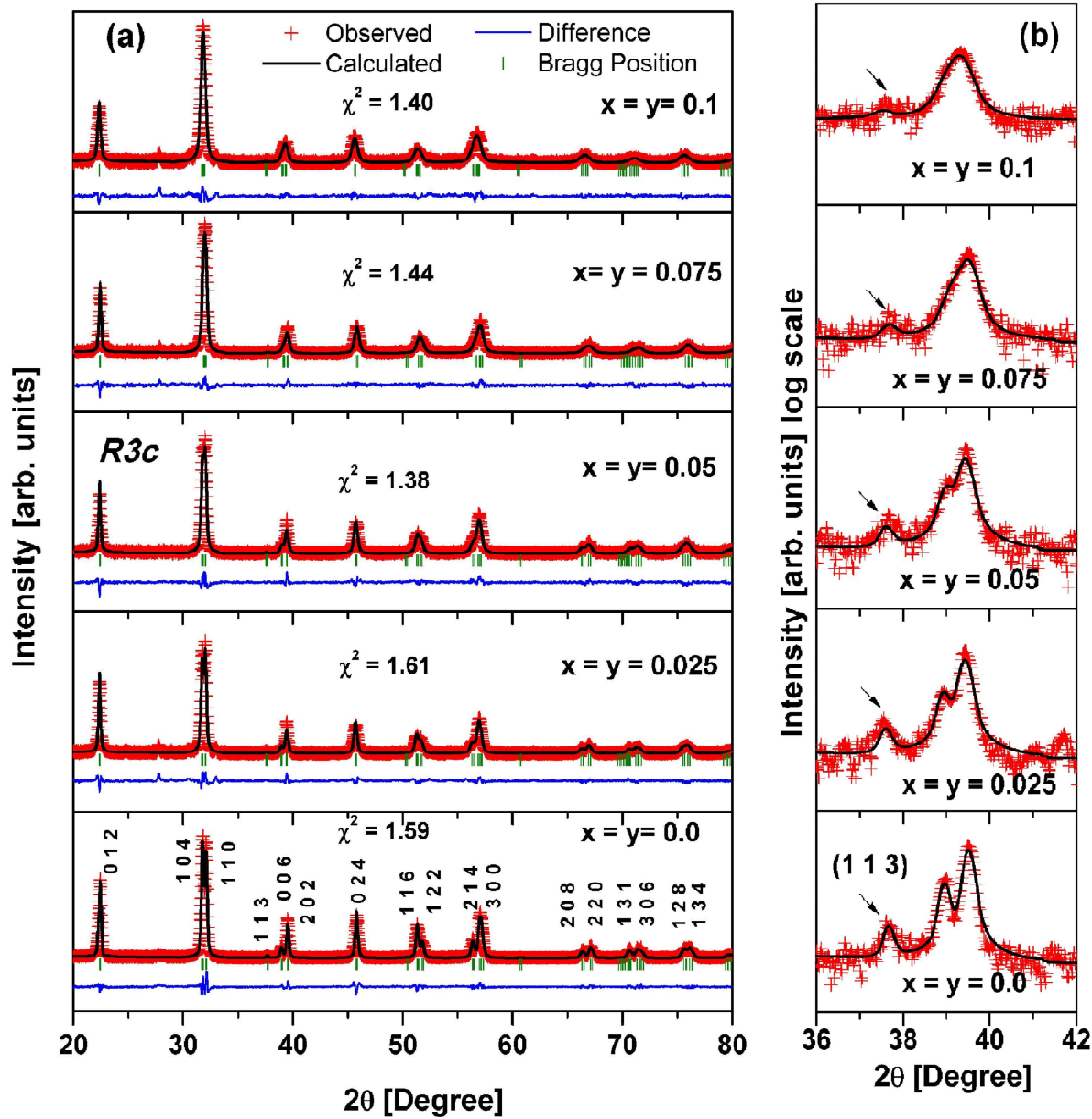
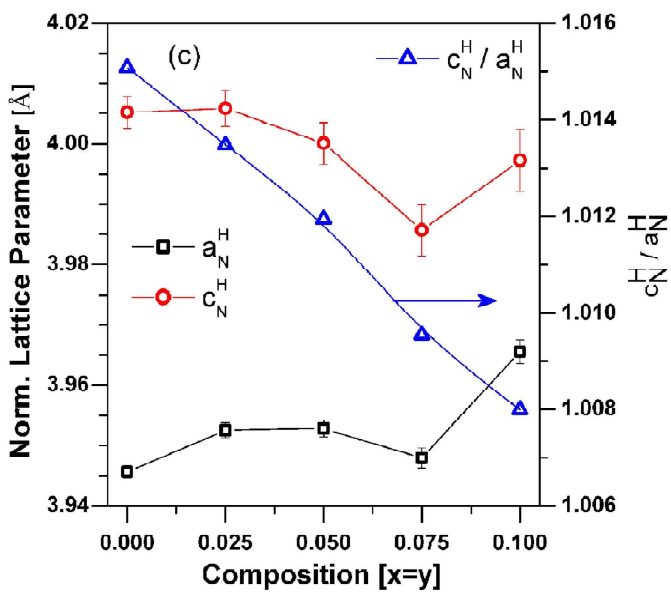
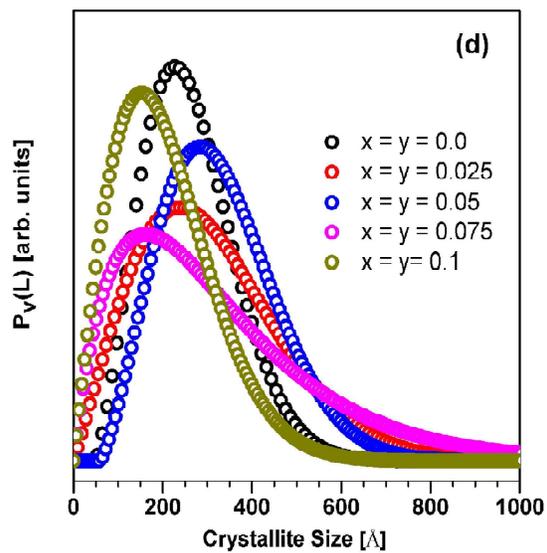

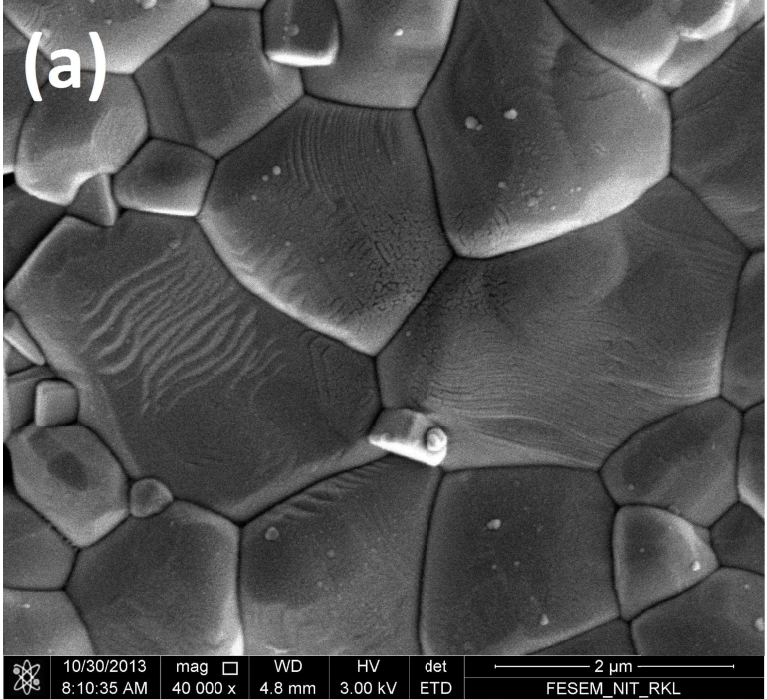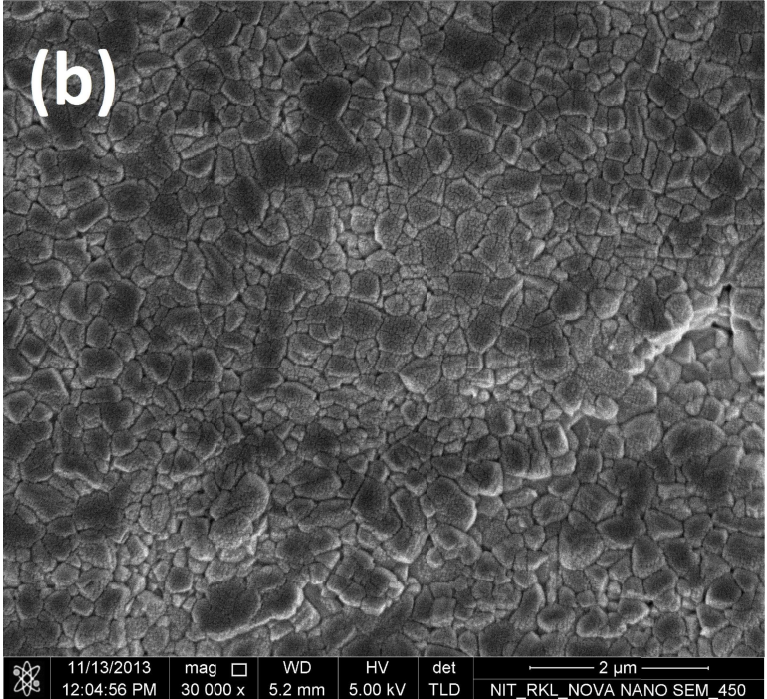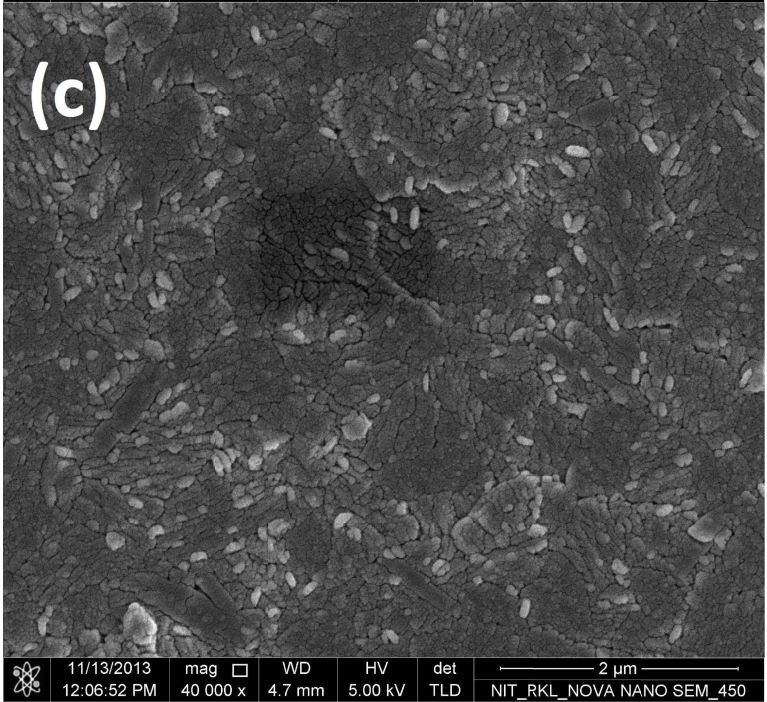

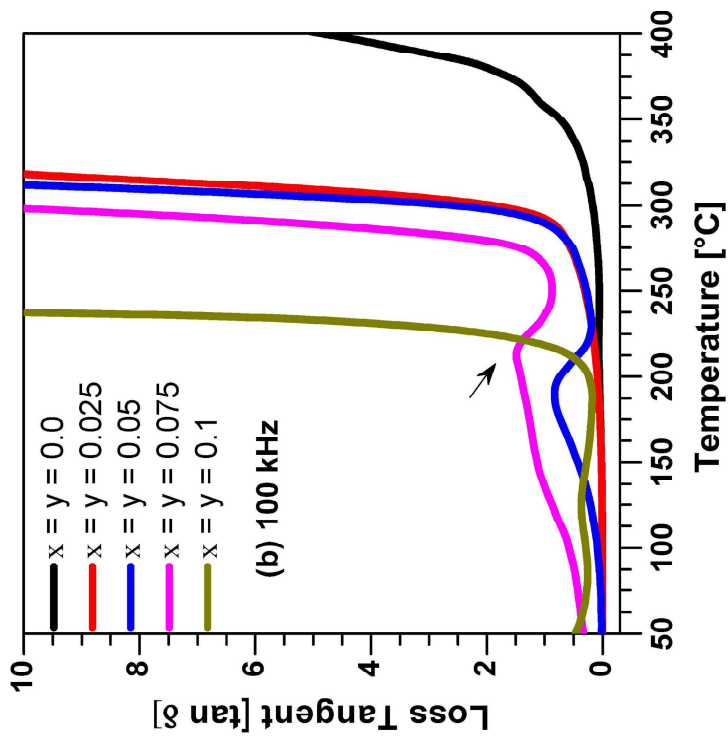
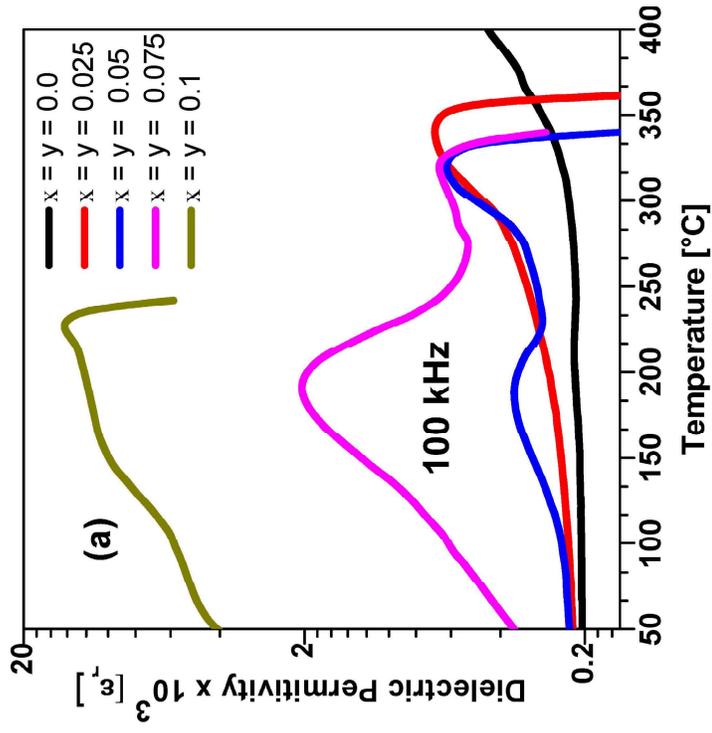
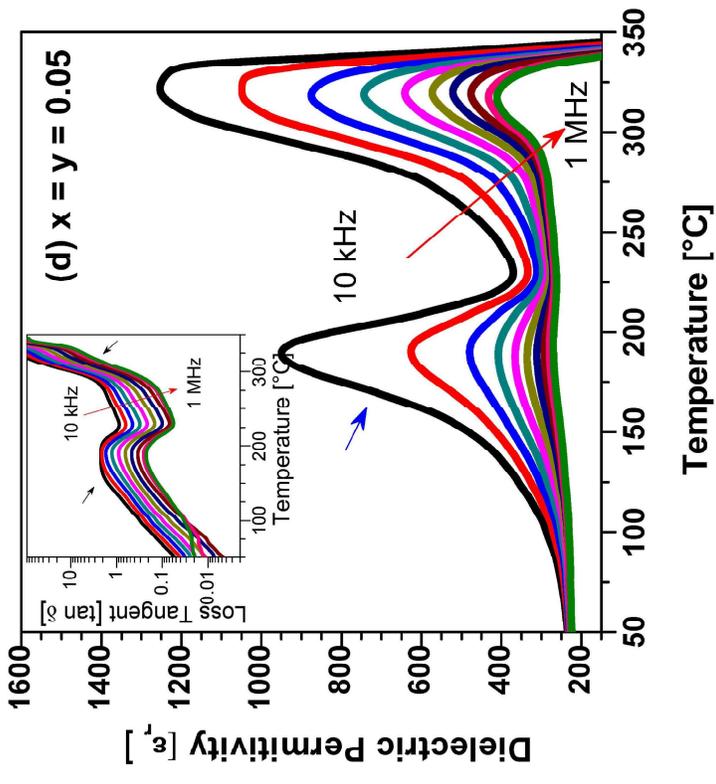
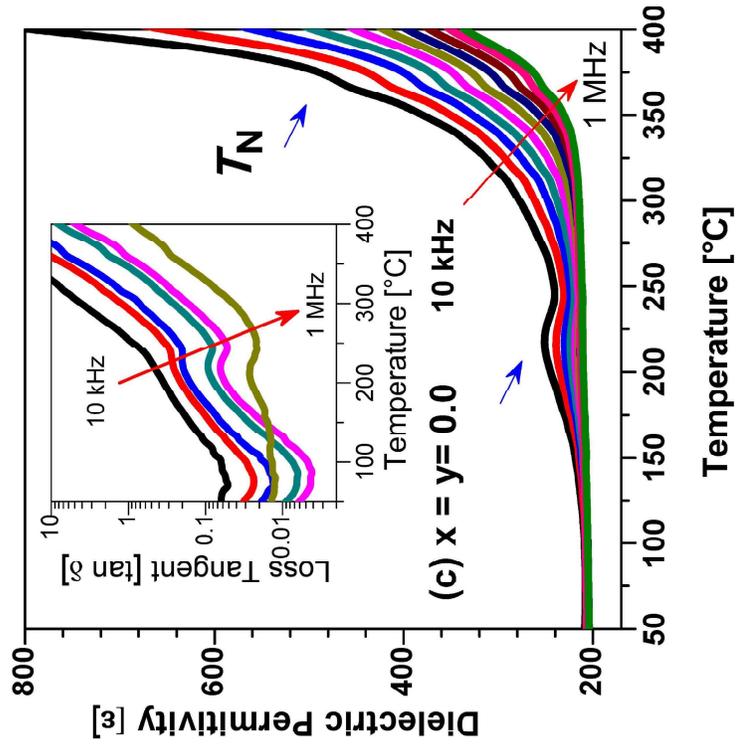

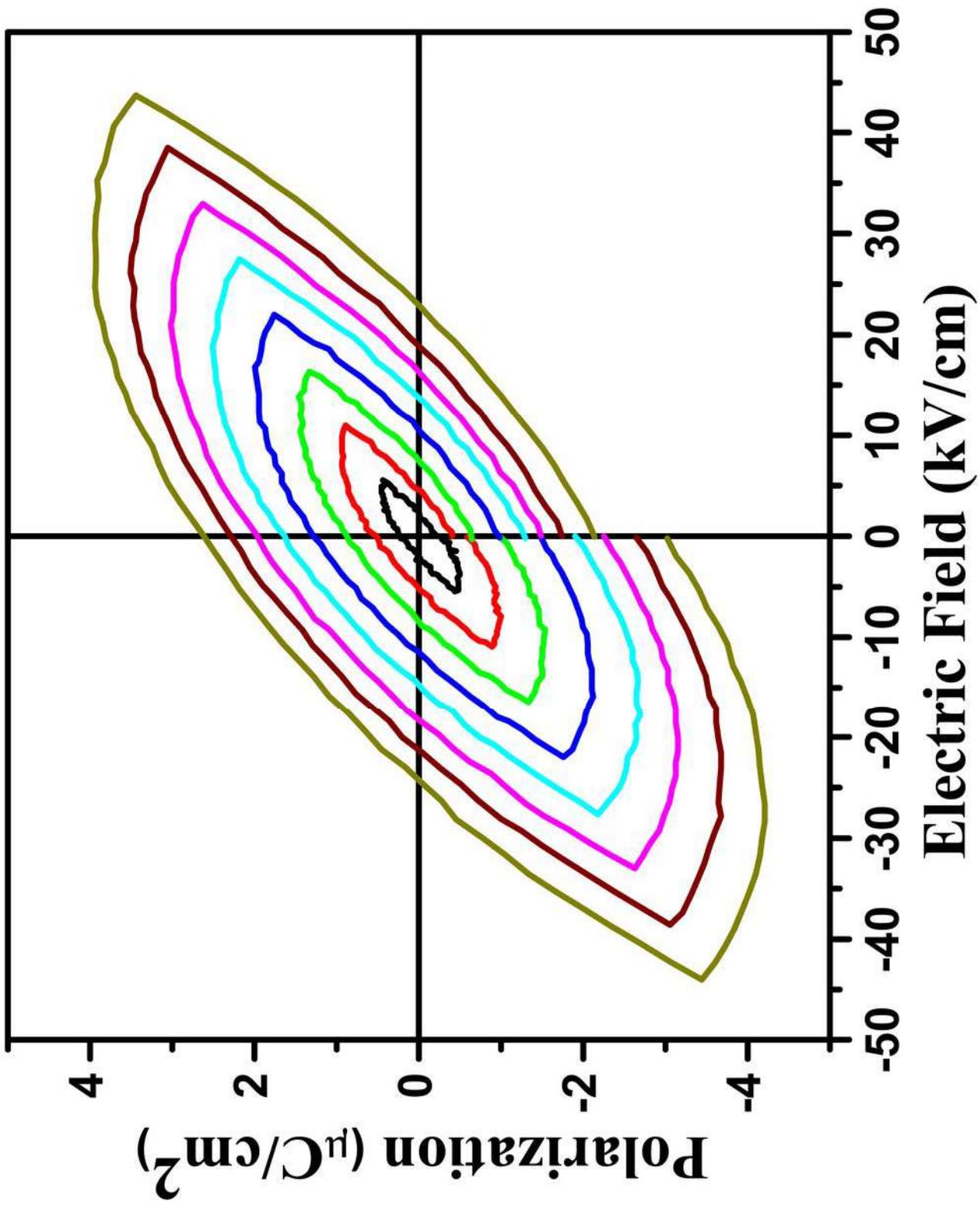

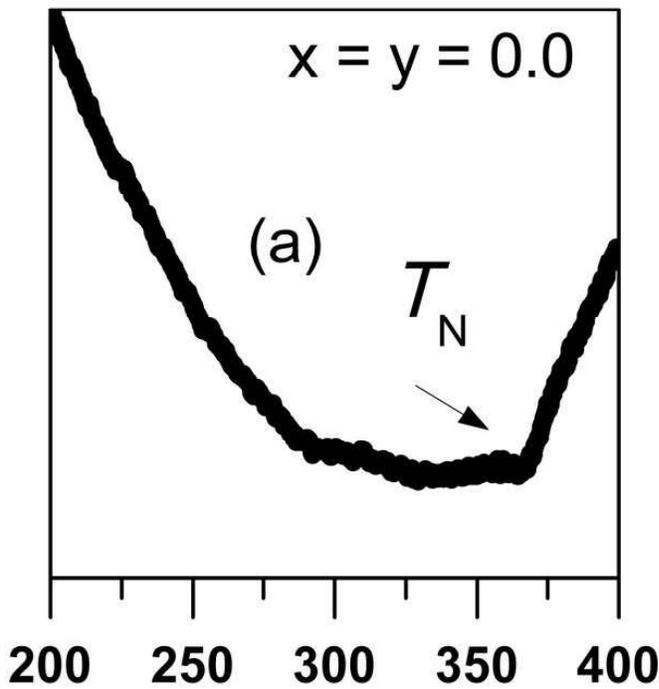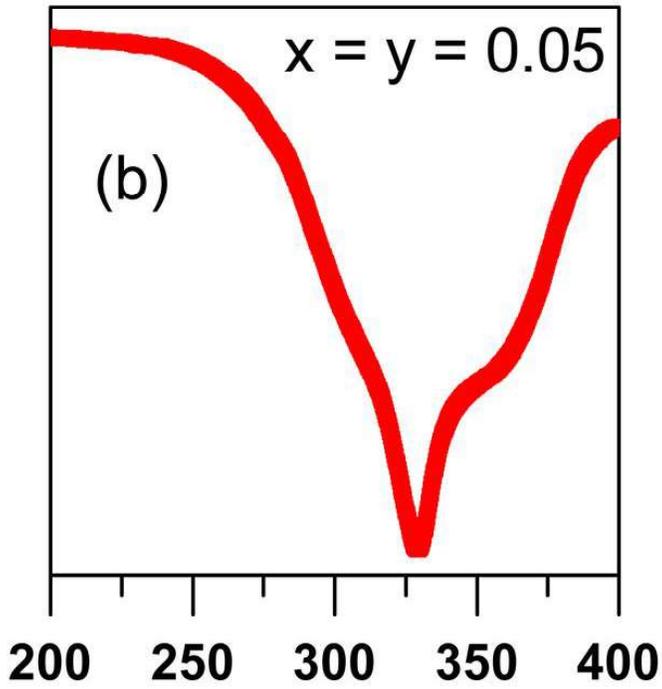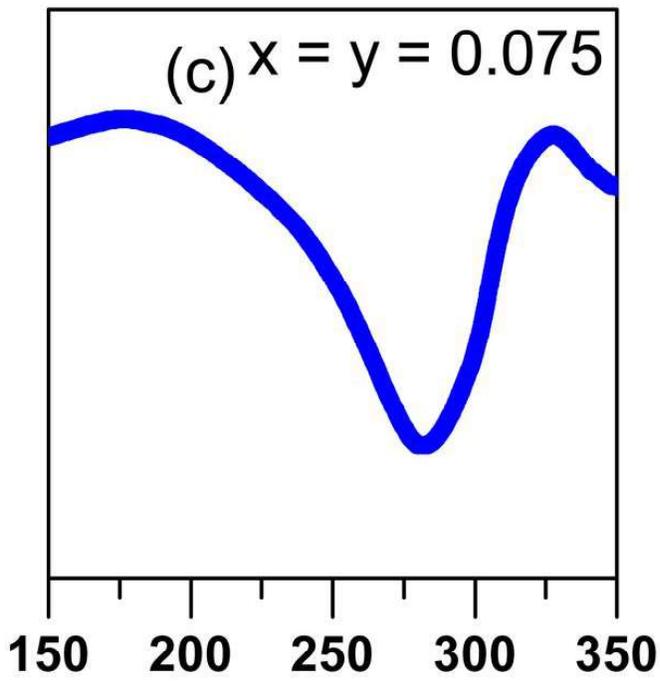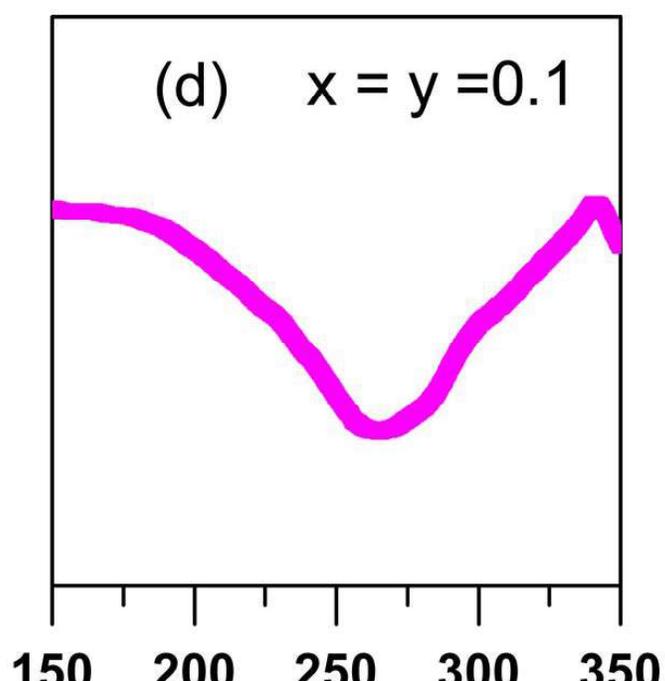

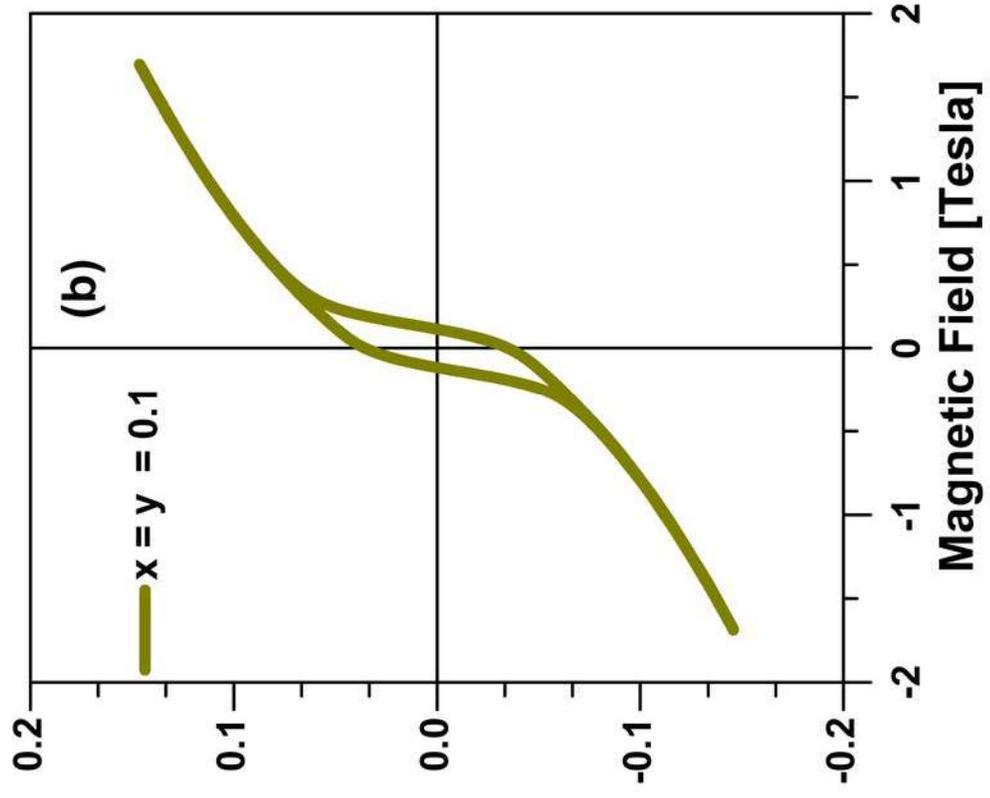
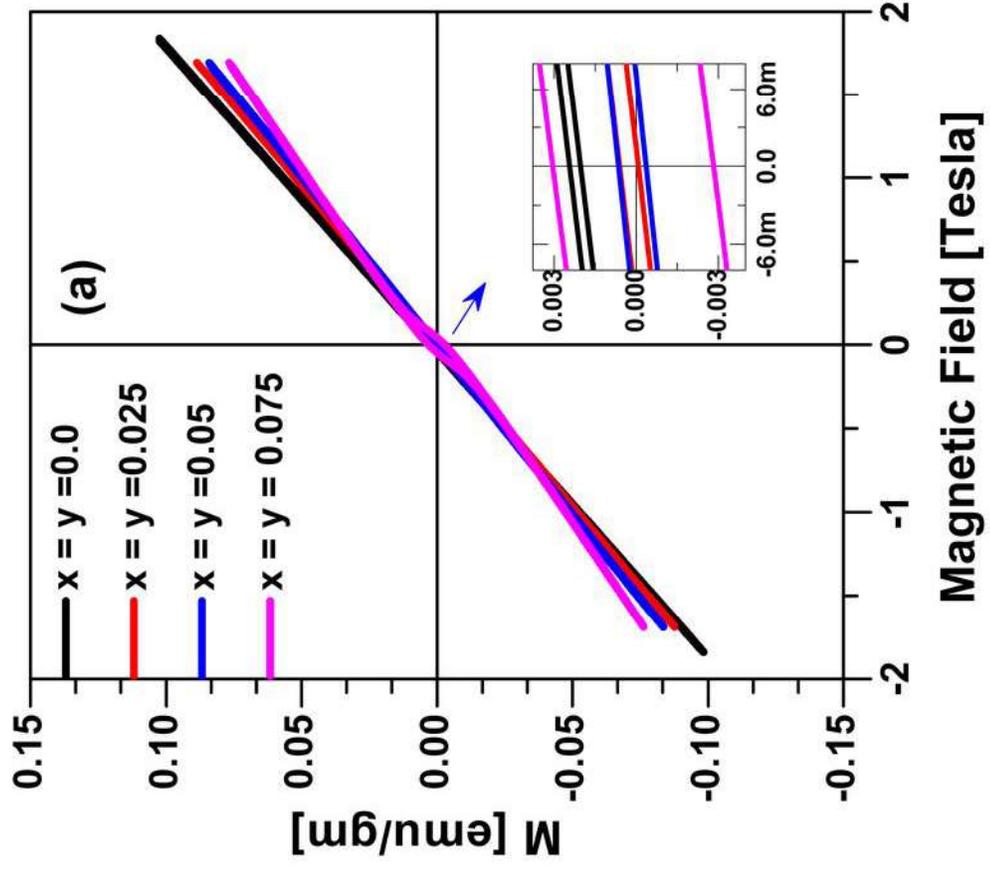

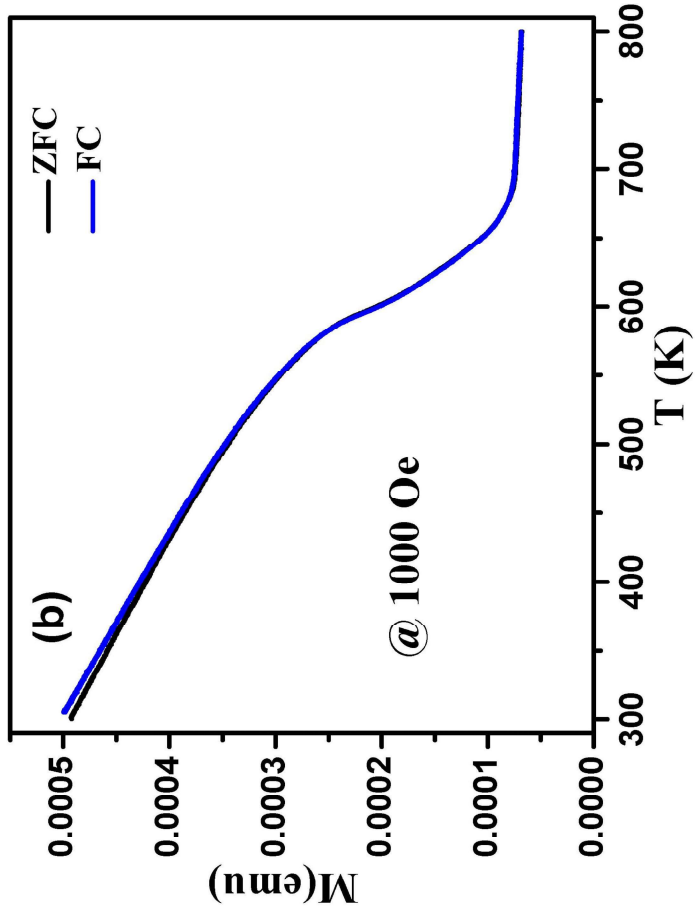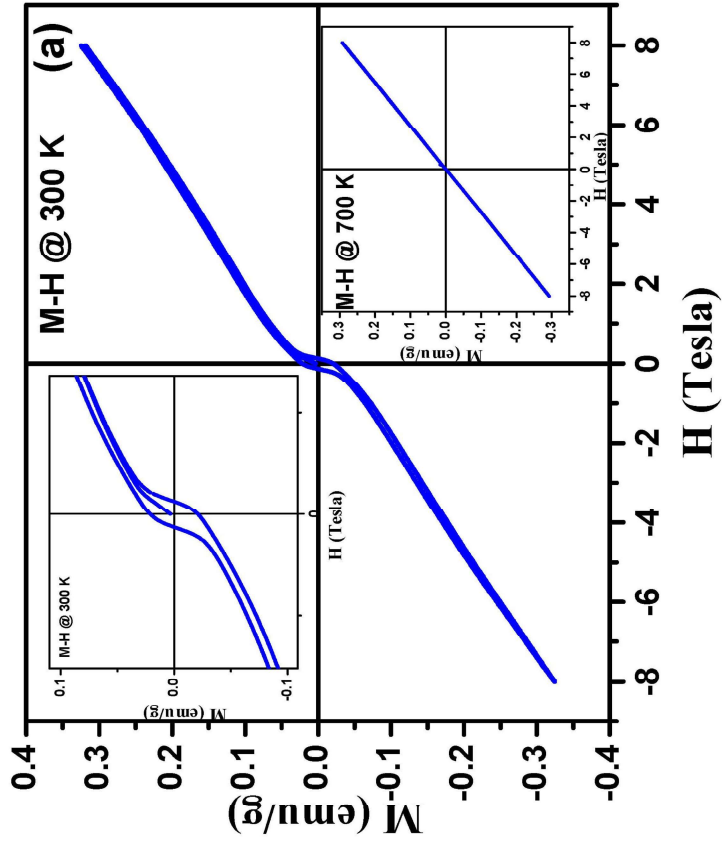

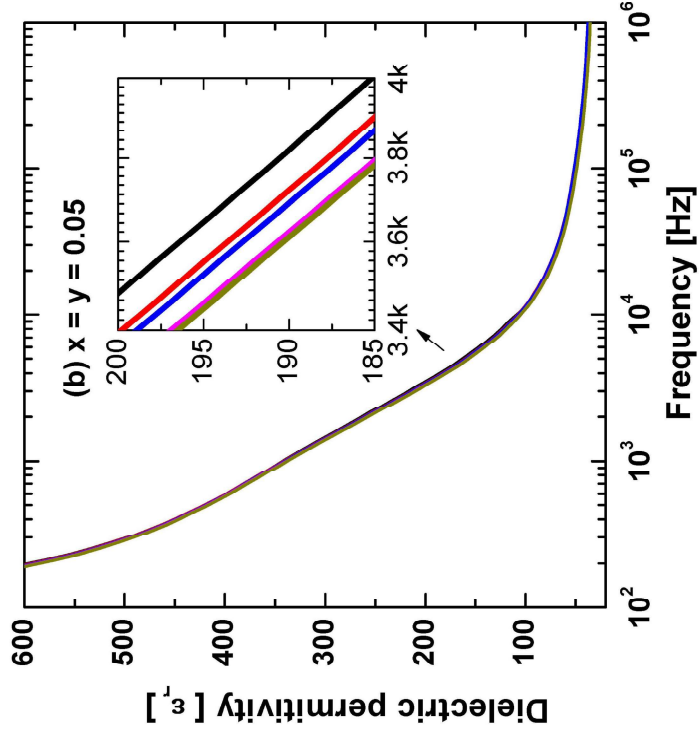
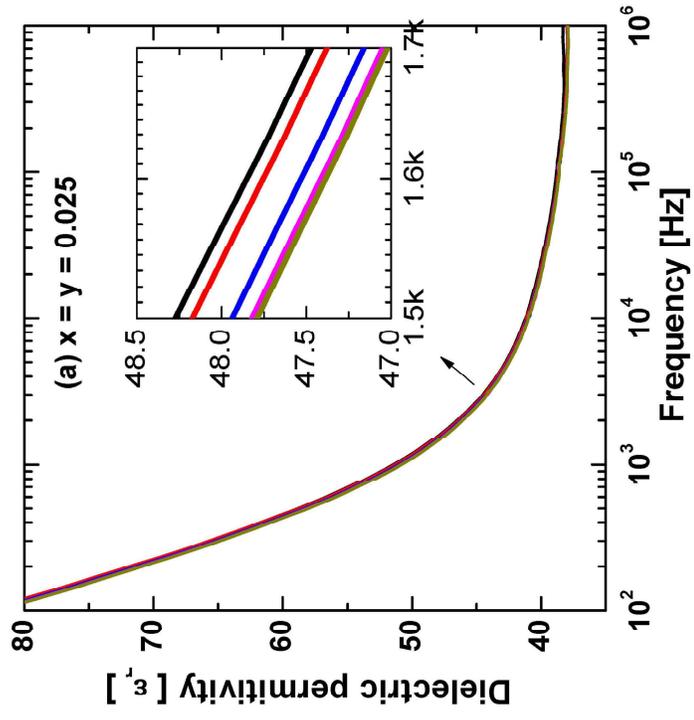
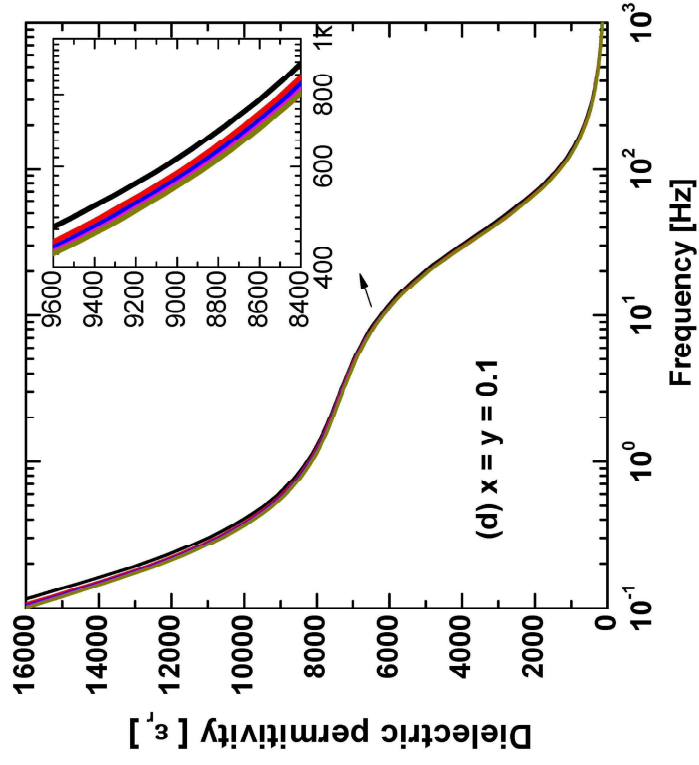
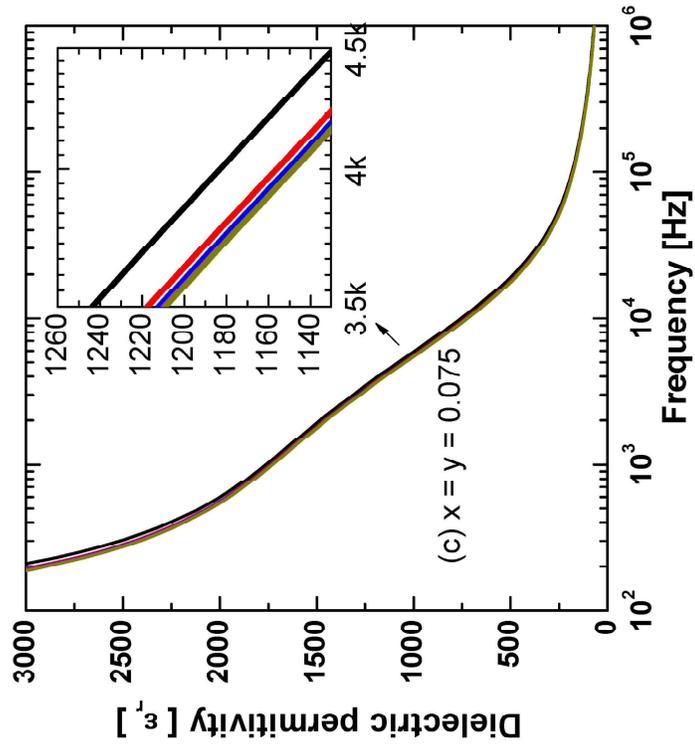

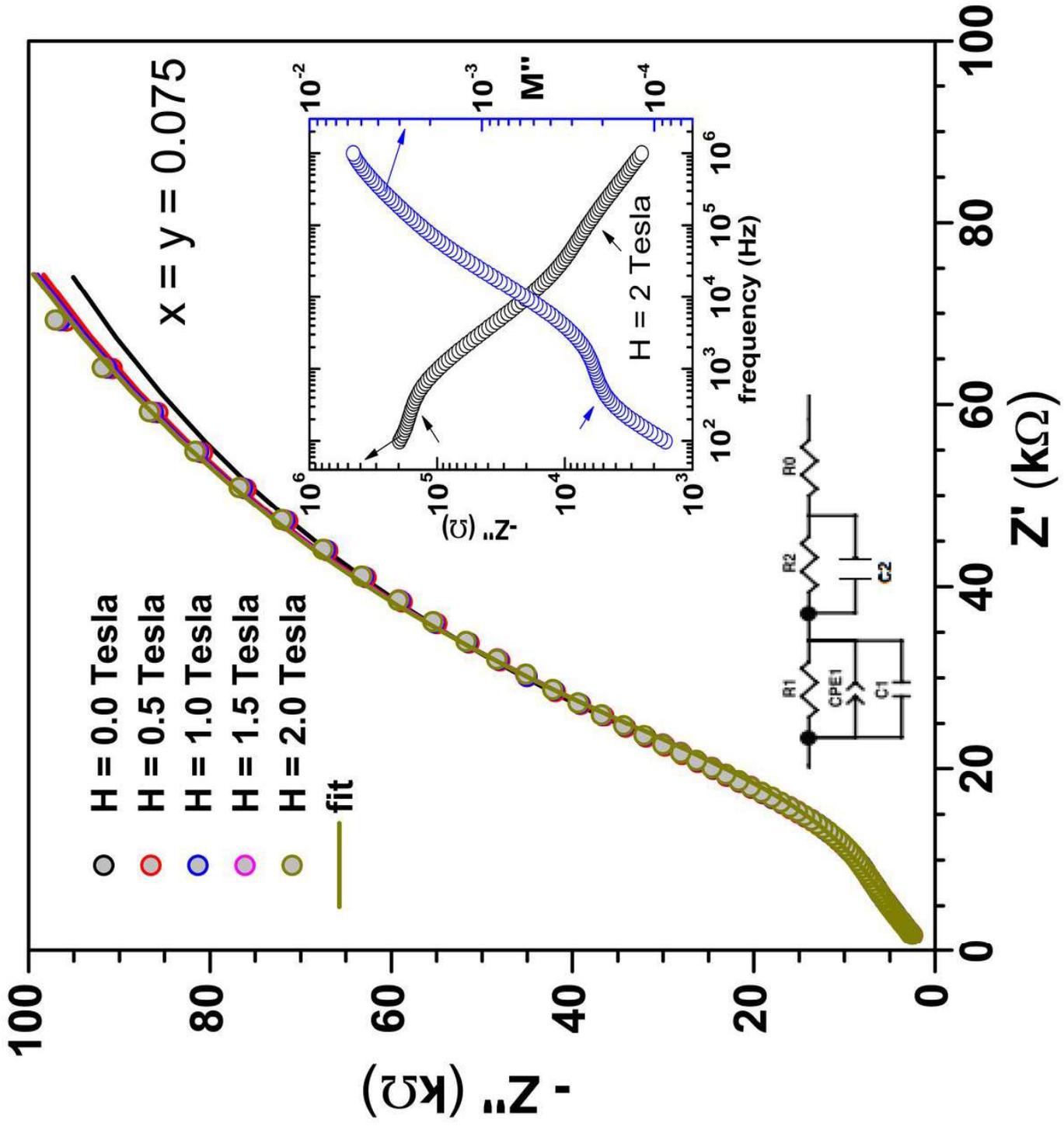